\documentclass[aps,twocolumn,showpacs]{revtex4}
\usepackage{graphicx}

\begin{document}
\draft

\newcommand{\beq}{\begin{equation}}
\newcommand{\eeq}{\end{equation}}
\newcommand{\ben}{\begin{eqnarray}}
\newcommand{\een}{\end{eqnarray}}
\newcommand{\bea}{\begin{array}}
\newcommand{\eea}{\end{array}}
\newcommand{\om}{(\omega )}
\newcommand{\bef}{\begin{figure}}
\newcommand{\eef}{\end{figure}}
\newcommand{\leg}[1]{\caption{\protect\rm{\protect\footnotesize{#1}}}}

\newcommand{\ew}[1]{\langle{#1}\rangle}
\newcommand{\be}[1]{\mid\!{#1}\!\mid}
\newcommand{\no}{\nonumber}
\newcommand{\etal}{{\em et~al }}
\newcommand{\geff}{g_{\mbox{\it{\scriptsize{eff}}}}}
\newcommand{\da}[1]{{#1}^\dagger}
\newcommand{\cf}{{\it cf.\/}\ }
\newcommand{\ie}{{\it i.e.\/}\ }

\title{\center{Nonclassical radiation from diamond nanocrystals}}

\author{Alexios Beveratos, Rosa Brouri, Thierry Gacoin$^*$,
Jean-Philippe Poizat and Philippe Grangier}
\affiliation{Laboratoire Charles Fabry de l'Institut  d'Optique, UMR 8501 du
CNRS, \\
B.P. 147, F-91403 Orsay Cedex - France \\
$^*$ Laboratoire de Physique de la mati\`ere condens\'ee,
Ecole polytechnique, F-91128 Palaiseau, France}

%
%

\begin{abstract}

The quantum properties of the fluorescence light
emitted by diamond nanocrystals containing a single
nitrogen-vacancy (NV) colored center are investigated.
We have observed photon antibunching with very low background light.
This  system is therefore a very good candidate for the production of
single photon on demand.
In addition, we have measured larger NV center lifetime
 in  nanocrystals than in the bulk, in good agreement with a simple
quantum electrodynamical model.

\end{abstract}


\maketitle

Light sources able to emit individual photons on demand would be of great
potential use for quantum cryptography \cite{TRG,L}.
A quantum computation scheme requiring such sources has also  been proposed
recently \cite{KLM}.
Considerable activity is thus dedicated to designing and implementing
efficient, robust, room-temperature sources delivering
a periodic train of pulses containing one and only one photon.
These sources are based on the property of a single emitting dipole to emit
only one photon at a time. When excited by a short and intense pulse, such
an emitter
delivers one and only one photon \cite{MGM,pra}.
 After pioneering experiments demonstrating
photon antibunching \cite{KDM,DW,BMOT} and conditional
preparation of single-photon states \cite{GRA,OM}, followed by
first attempts to build triggered single photon sources \cite{MGM,BLTO,KBKY},
the present generation of experiments is concentrating on
solid-state schemes better suited for practical use, such as
single organic molecules \cite{KJRT,wild,ML,TCGR},
 self-assembled semiconductor quantum dots \cite{YY,I2}, or
semiconductor nanocrystals \cite{I1}. The succesful candidate
should work at room temperature, and be photostable.

A promising system for a robust single photon source is provided by
individual nitrogen-vacancy (NV) color centers
in diamond \cite{GDTFWB}, that already permitted  to observe photon
antibunching
under continuous excitation conditions in bulk crystals \cite{KMZW,ol,capri}.
These color centers have the great advantage of being  photostable and do not
exhibit any photoblinking.
The set-up in these experiments is particularly simple,
since it involves diamond samples at room temperature,
and non-resonant excitation from a laser at  $532$ nm,
with a typical power in the mW range.
However, a significant limitation in bulk diamond is that the light is emitted
in a high-index material ($n_{d}=2.4$), that makes its efficient extraction
difficult.
 Refraction at the sample interface leads to a small collection
solid angle and to aberrations.
A similar problem arises in semiconductor light-emitting devices \cite{wb}.

In this paper, we show that diamond nanocrystals (typical size $40$ nm)
containing a single NV center
offers a very efficient solution to circumvent these problems\cite{KHSPV}.
The subwavelength size of these nanocrystals renders refraction irrelevant.
One can  simply think of the nanocrystal as a point source emitting light
in air.
Furthermore, the small volume of diamond excited by the pump light yields
very small background light. This is of crucial importance
for single photon sources, since background light contributes to a
non-vanishing
probability of having two or more photons within the light pulse.

In addition, the width of the dip of the fluorescence
intensity autocorrelation function $g^{(2)}(t)$
gives information about the lifetime of the emitters. Using this technique,
we have observed an  increase  of the lifetime
of a NV center in a nanocrystal compared to the bulk value \cite{I3}.
This effect arises from the fact that,
in a nanocrystal,   the center can be considered as radiating in air,
whereas it radiates in a
medium of refractive index  $n_d=2.4$ in bulk \cite{na}.
We will
make the assumption, that is consistent with our
observations, that  the local field
 experienced by the NV center
is the same in the bulk and in a nanocrystal. Our observed change
in lifetime would thus be independent of local field corrections, that have
been a controversial  issue during the last decade \cite{gl,cc,qph}.

The color center used in our experiments is the NV defect center
in synthetic diamond, with a zero phonon line at a wavelength of $637$nm
\cite{GDTFWB}.
The defect consists in a substitutionnal nitrogen and a vacancy in a adjacent
site. A simplified level structure is a four-level scheme
with fast non radiative decays within the two upper states
and within the  two lower states. The excited state
lifetime  in the bulk is $\tau_b =11.6$ ns \cite{CTJ}.
The  nanocrystals come from synthetic diamond powder  bought from de Beers.
The defects are created  by  irradiation with $1.5$ MeV electrons at a dose
of $3\times 10^{17} e^{-}/$cm$^2$, and annealing in vacuum
at $850^o$C during $2$ hours. The density of NV centers created is then
estimated to be of  one  in a $30$ nm diameter sphere \cite{GDTFWB}.
The nanocrystals are  dispersed by sonification in a solution of polymer
(Polyvinylpyrrolidone at 1 weight$\%$ in propanol). This allows the
disaggregation of the particles and their stabilisation in a colloidal
state. Centrifugation
 at 11000 rpm for 30 mn allows us to select nanocrystal sizes of $d_0=90\pm
30$ nm (measured by dynamical light scattering).
The average number of NV centers in a nanocrystal has been evaluated to $8$
(see below).
Nanocrystals containing a single NV center should
therefore have a size around $d_0 /2 =45$ nm, that lies
in the lower wing of the size distribution.  The nanocrystal
solution is then  spin coated at 3000 rpm on thin fused silica
substrates. Evaporation of the solvent leaves a 30 nm thick film of polymer
with the nanocrystals well dispersed on the surface. Their density was
estimated to be around  $0.02$ $\mu$m$ ^{-2}$.
In most experiments we look at the centers from the other side of the
plate, that is in contact with the oil of an immersion microscope lens
(Nachet 004279, N.A. = 1.3).

\begin{figure} [!ht]
\includegraphics[scale=0.3]{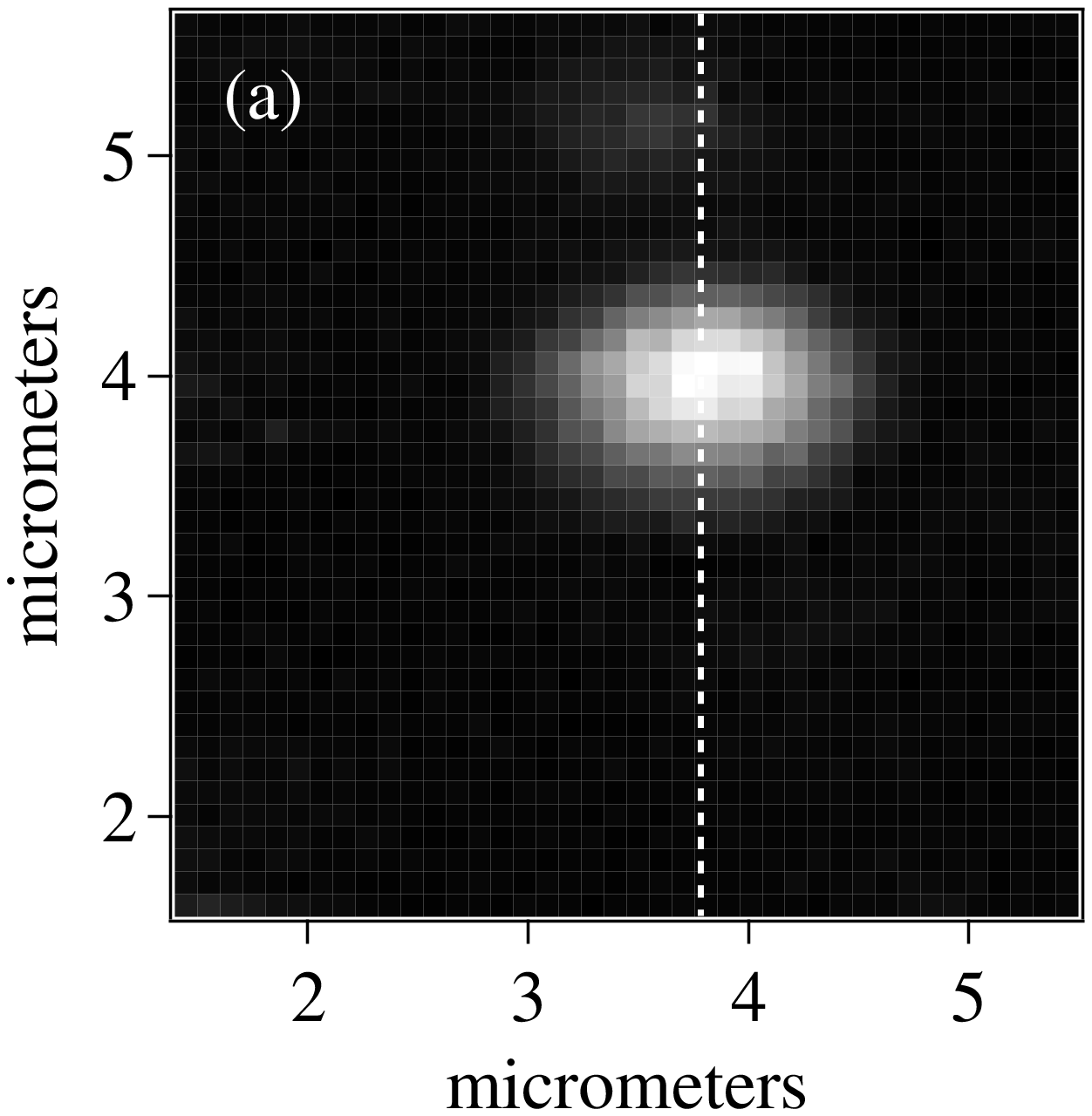}
\includegraphics[scale=0.3]{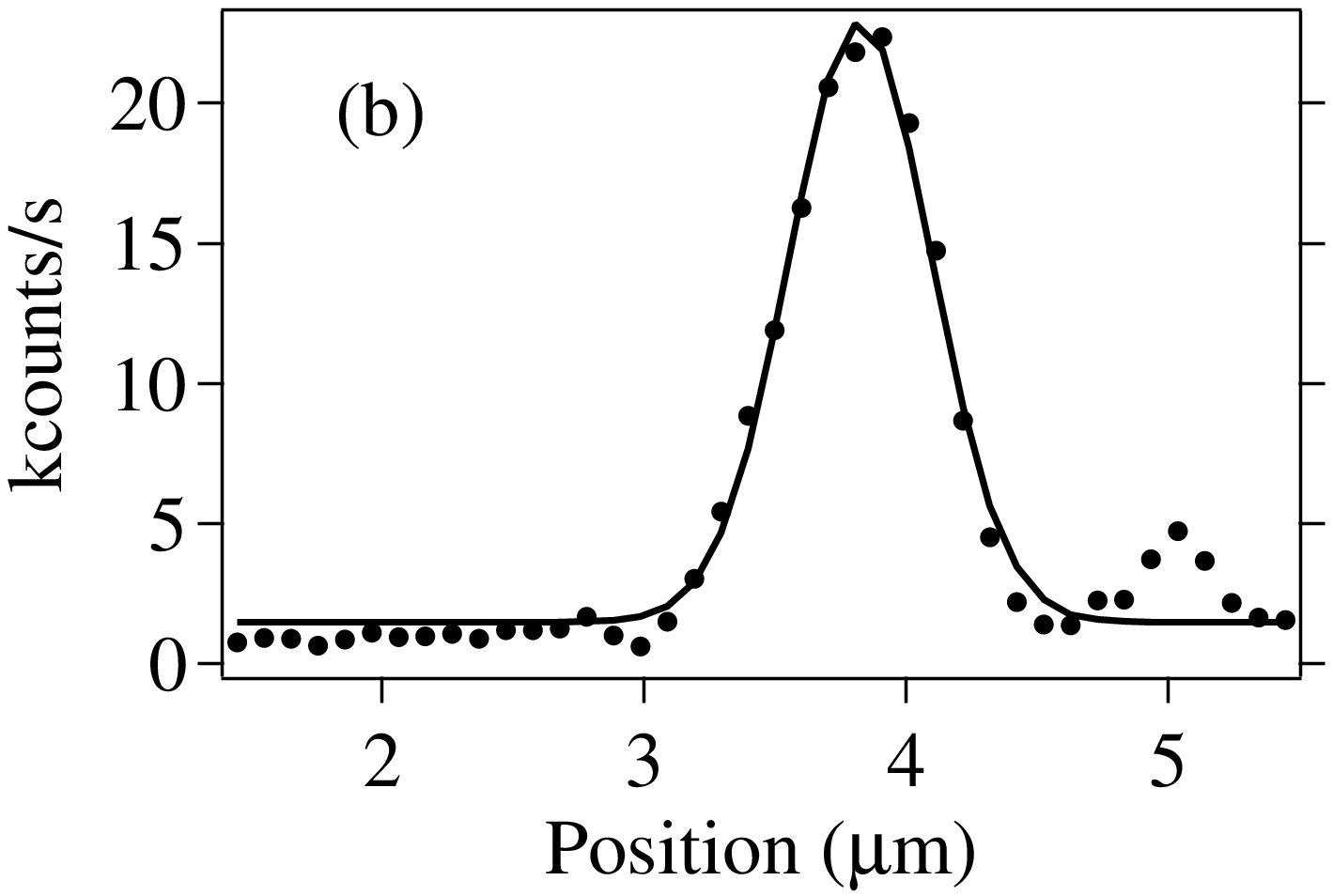}
\caption{(a) Confocal microscopy raster scan
($ 5 \; \times \;  5 \; \mu$m$^2$) of a diamond nanocrystal containing a
single NV center. The count rate corresponds to one detector only.
The size of a pixel is $100$ nm and the integration time per pixel is $32$ ms.
The laser intensity impinging on the sample is $2.7$ mW.
In (b) is displayed a linescan along the dotted line of (a), together with
a gaussian fit,
that is used to evaluate the signal and background levels.
Here we obtain $ S/B=20$.}
\label{scan}
\end{figure}

The experimental set-up has been described in detail elsewhere \cite{ol}.
It is based upon a home-made scanning confocal microscope,
where the sample is excited using continuous-wave frequency-doubled YAG laser
($\lambda = 532$ nm).
The fluorescence light (wavelength between $637$ and $800$ nm)
is collected by the same objective
and separated from the excitation laser by a dichroic mirror and filters.
It can be sent either to a spectrometer, or to a
standard intensity correlation set-up using two
avalanche photodiodes (EG$\&$G, model SPCM-AQR 13), a
time to amplitude converter (TAC) and  a computer data acquisition
board. The time bin is 1 ns, and a delay of $50$ ns
in one TAC input allows us to get data for negative time.
A slow (8 s response time) x-y-z computerized servo-lock is used to maintain
the fluorescence on its maximum for the observed center.

We have checked that the emission spectrum of a NV center in a nanocrystal
at room temperature is identical to that in bulk to  within 
our experimental  accuracy (resolution $8$ nm,
signal to noise ratio larger than $10$).
It is also worth pointing out that the remarkable photostability of NV
centers in bulk \cite{GDTFWB,KMZW,ol} is preserved in nanocrystals.
Fluorescence has been observed in the saturation regime for hours without
 any photobleaching nor blinking.

Fig. \ref{scan}(a) displays a 2D scan of a nanocrystal containing a single NV
center.
The resolution of $500$ nm is that of the confocal microscope.
The line scan in Fig. \ref{scan}(b) shows that the  signal ($S$) to
background ($B$)  ratio  is
very good with a value $S/B= 20$.
Note that $B$ is the count rate measured about 2 $\mu$m away from a nanograin.

\begin{figure}[!ht]
\includegraphics[scale=0.38]{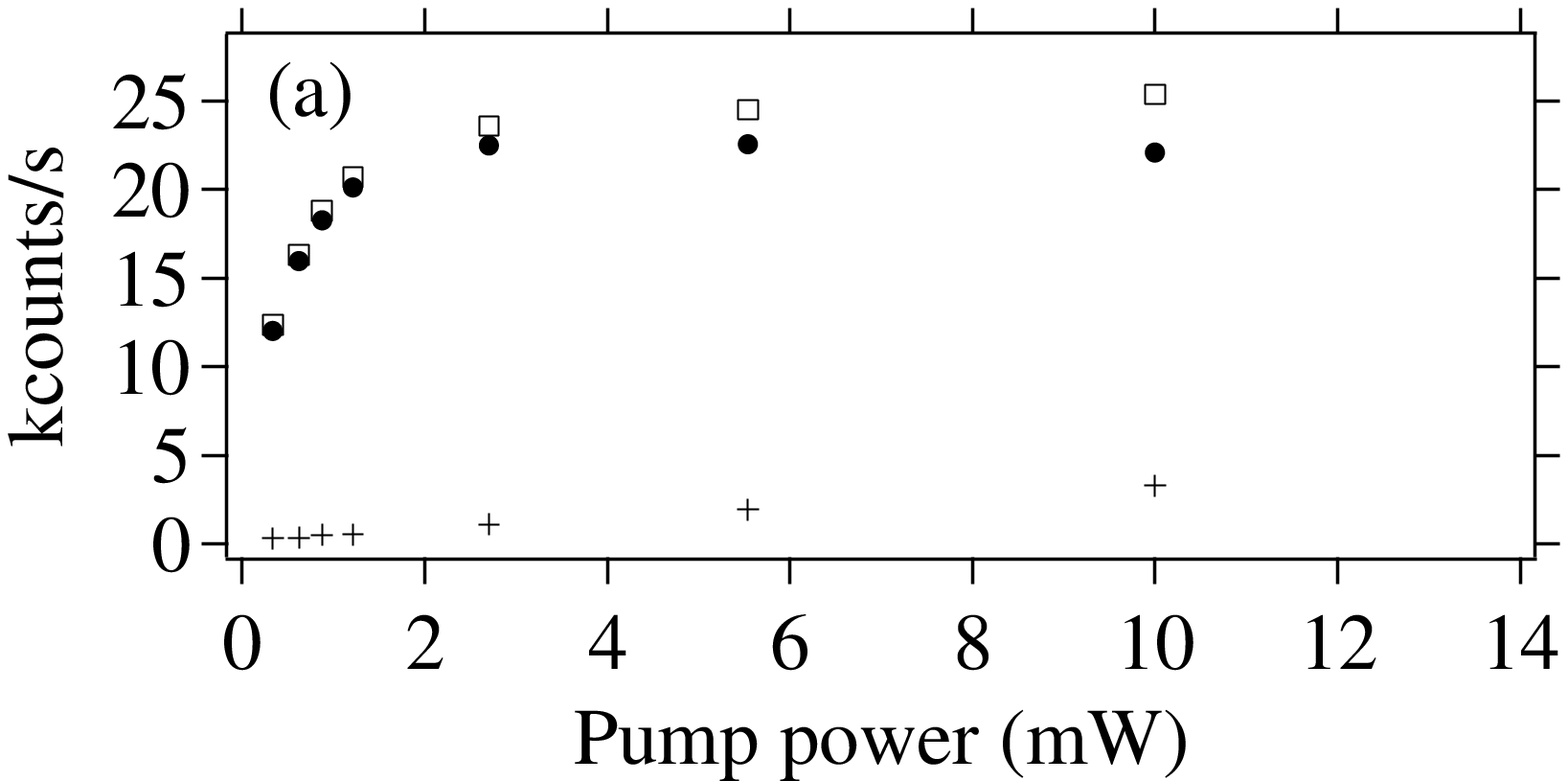}
\includegraphics[scale=0.38]{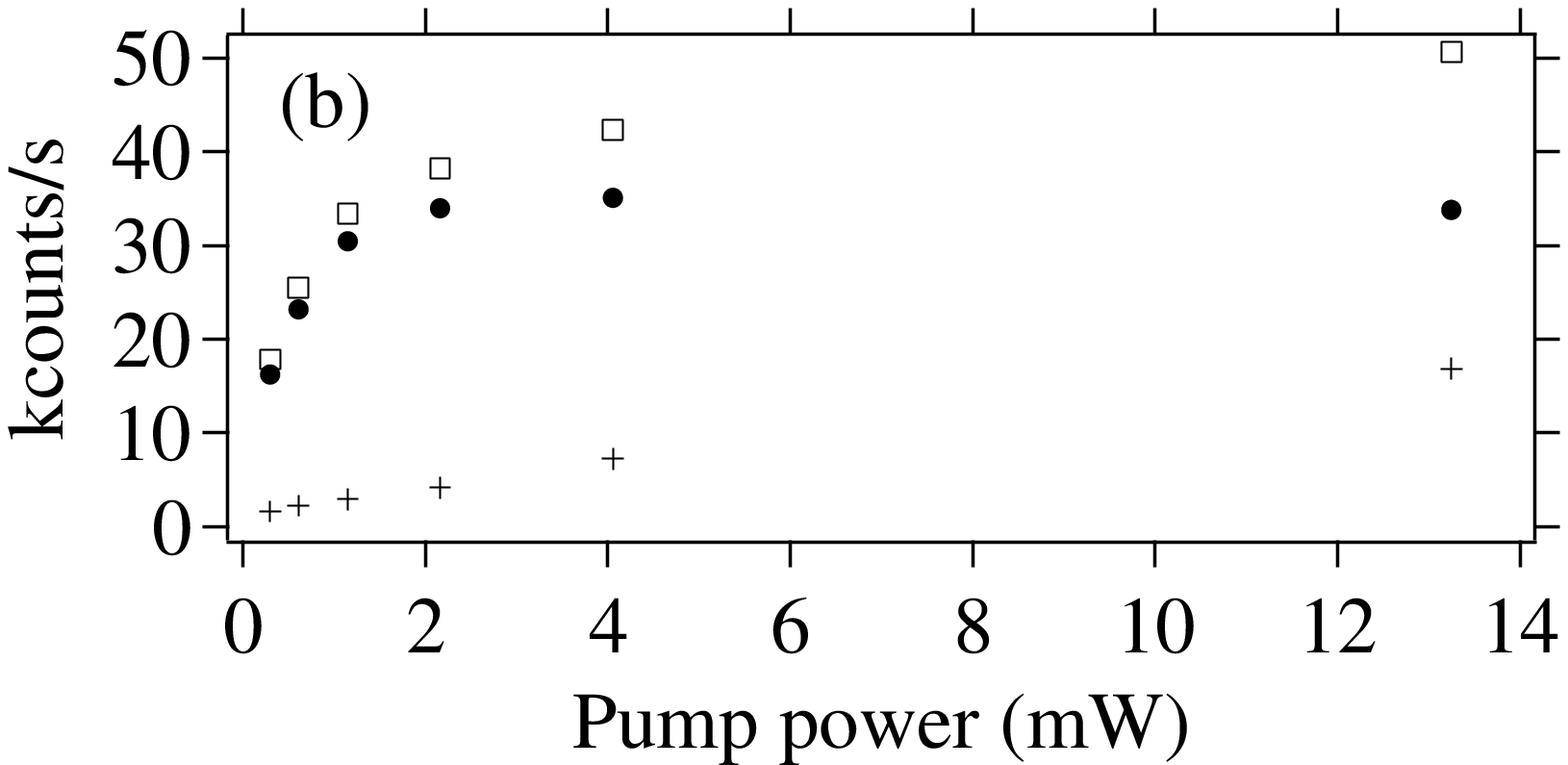}
\caption{Fluorescence rate of a NV center in a nanocrystal (a) and in bulk
diamond (b)
as a function of the pump power.
The count rate corresponds to one detector only.
The crosses, empty squares, and black circles represent
 respectively the background $B$, the total count rate $T=S+B$,
and the signal from the center $S=T-B$.
The maximum number of photons emitted in a lifetime is
$5.5 \times 10^{-4}$ for the nanocrystal (lifetime $\tau_{nc}= 25$ ns)
and $3.7 \times 10^{-4}$ in the bulk (lifetime $\tau_{nc}= 11.6$ ns).
The data for the nanocrystal corresponds to the center  of Fig.\protect\ref{scan}.}
\label{fsat}
\end{figure}

Fig. \ref{fsat} shows the fluorescence rate of a NV center in a nanocrystal
and in bulk diamond
as a function of the pump power.
Slightly decreasing rate for high pump power is attributed to the presence
of an additionnal
shelving state \cite{capri,DFTJKNW}. 
It can be seen that the contribution of the background is
greatly
reduced in the nanocrystal configuration.
The count rate in the nanocrystal is not as high as expected.
However a fair comparison with the bulk
should take into account the $\tau_{nc}/\tau_b=2.2$
factor increase of the NV center lifetime
in a nanocrystal (see below).
The number of photons emitted in a lifetime is then larger in the nanocrystal.

\begin{figure} [!ht]
\includegraphics[scale=0.4]{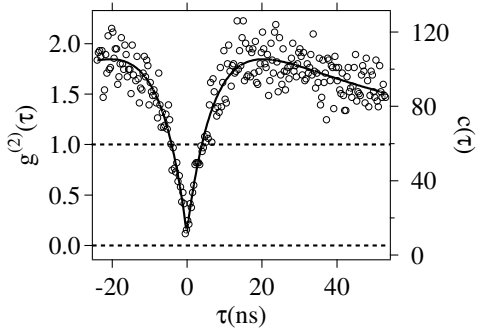}
\caption{Autocorrelation function $g^{(2)}(t)$ (left) and raw coincidence
rate (right).
The time bin is $w=0.3$ ns,
the total integration time is $323$ s, and
the laser intensity impinging on the sample is $2.7$ mW.
Count rates on each 
photodiode are $N_1=22500$ s$^{-1}$ and $N_2=24500$ s$^{-1}$.
The actual number of coincidences is indicated on the right.
The zero-time value of the uncorrected normalized
correlation function is $C_N(0) = 0.17$.
The fit is performed with the model used in \protect\cite{capri}.
The data corresponds to the center  of Fig. \protect\ref{scan}. }
\label{ABmc}
\end{figure}

The raw coincidences $c(t)$ (right axis) and  autocorrelation function
$g^{(2)} (\tau ) = \langle I(t)I(t+ \tau ) \rangle / \langle I(t) \rangle ^2$
(left axis) are represented in Fig. \ref{ABmc}.
For evaluating  the  intensity correlation function $g^{(2)}(t)$  of the NV
center,
the raw correlation data $c(t)$ is normalized and corrected  in the
following way.
The raw coincidence rate $c(t)$ counted during a time $T$ within a time bin
of width $w$
is first normalized to that of a Poissonian source according
to the formula $C_N(\tau ) = c(\tau) /(N_1 N_2 w T)$,
where $N_{1,2}$ are the count rates on each detector.
Then the normalized coincidence rate $C_N(\tau ) $ is corrected for
the background light, and we obtain
$ g_c^{(2)}(\tau ) = [C_N(\tau ) - (1 - \rho ^2)]/\rho^2$, where
$\rho = S/(S+B)$ is related to the signal to background ratio,
that is measured independently in each experimental run.
Note that we have checked experimentally that the
background light has a Poissonian statistics.
It can be seen in Fig. \ref{ABmc} that $g^{(2)}(0 ) \sim  0$, where
the slight difference with
zero is attributed to remaining background light emitted by
the nanograin.
This almost vanishing value of $g^{(2)}(0 )$ is
the signature of the presence of  a single emitter in the observed nanocrystal.
In the case of the presence of $p$ centers within an unresolved peak,
the value of the zero-time antibunching is $1-1/p$. This is actually how we
estimate
the number of NV centers in a nanocrystal.
It should  also be mentionned that $g^{(2)}(\tau )$ reaches values greater
than unity
for $\tau \neq 0$.
This bunching effect is due to the presence of a third state in which the
system can be shelved
\cite{capri,DFTJKNW}.

Obviously, for the ultimate achievement of a true single
photon source, no background subtraction is possible and
the  meaningful quantity is the experimentally measured $C_N(0)$.
Indeed, when the center is excited by a short an intense pulse, the
 probability $p_2$ of having more than 2 photons in
a pulse is given by (assuming $p_2 \ll 1$)
\beq
p_2= C_N(0) \; p_1^2/2
\label{p2}
\eeq
where $p_1$  is the probability of having a single photon.
Note that for Poissonian light $C_N(0)=1$ and eq.(\ref{p2}) gives the
photon probability distribution of an attenuated coherent pulse.
In our case $C_N(0)=0.17$ at the fluorescence rate maximum (input power of
$2.7$ mW),
 where the best value in bulk was $0.26$  \cite{KMZW,capri}.
This would yield to
significant improvement compared to
attenuated coherent pulses that are usually used in quantum cryptography
experiments \cite{TRG}.

\begin{figure} [!ht]
\includegraphics[scale=0.4]{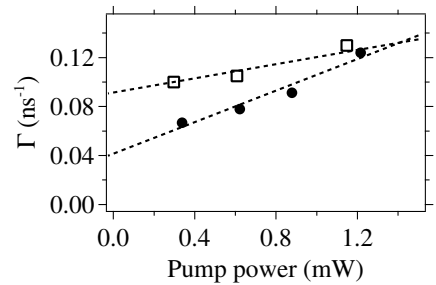}
\caption{Width of the exponential dip of antibunching traces obtained at
different pump power. The extrapolation for vanishing pump power gives
the lifetime. The black circles correspond to the data of
Fig. \protect\ref{scan}
and yield $\tau _{nc}= 25$ ns.
The empty squares correspond to a NV center in bulk diamond ($\tau_b=11.6$ ns).
The slope in the nanocrystal is twice as large as the one in bulk. This
is consistent with the lifetime increase, since the NV center in a nanocrystal
can absorb twice as many  photons during its lifetime.}
\label{ABtau}
\end{figure}

The central dip in the antibunching traces can be fitted by an exponential 
function of argument $-(r+\Gamma)\be{\tau}$, where   $r$ and $\Gamma$
are
respectively the pumping rate and the NV center decay rate.
Such fits have been performed for different pumping powers.
The inverse lifetime $\Gamma=1/\tau _{nc}$ of an NV center in a nanocrystal
can  then be inferred by  extrapolating the value of the time constant for
vanishing pump power.
Fig. \ref{ABtau} shows a lifetime of $\tau_{nc} = 25$ ns for the observed
NV center,
that  is significantly longer than the lifetime
of a NV center in synthetic bulk diamond,
$\tau_b = 11.6 \pm 0.1$ ns \cite{CTJ}.
This effect can be attributed to the strong change in the refractive index
of the surrounding medium, when going from bulk diamond to nanocrystals.
In a simple approach\cite{na}, the squared amplitude of the one-photon
electric field
should be divided by the relative susceptibility $\epsilon_r = n^2$,
while the mode density, that is proportional to the elementary volume
in the wavevector space, should be multiplied by $n^3$. Since the
inverse lifetime is proportional to the product of these two quantities,
one obtains the simple formula\cite{na}
\beq
\Gamma_n = n \Gamma_v
\label{na}
\eeq
relating the spontaneous emission rates $\Gamma_n$ in the material
and $\Gamma_v$ in the vacuum. However, this formula does not take into account
the local field effects due to the possible
 modification of the immediate surroundings of the emitting dipole,
and the correct formula turns out to be $\Gamma_n = n l^2 \Gamma_v$
where $l$ is the local field
enhancement factor. Different models leading to different
local field correction factors have been proposed,
and this topic is actively discussed in the
literature \cite{gl,cc,qph}.

However our experimental results agree well with  eq. (\ref{na}),
and thus seem insensitive to the local field correction $l$.
This may be explained by noticing that
the immediate environment of the NV center at a scale of $\sim 1 nm$
(ie a few crystalline periods) is unchanged.
Therefore,  the local field experienced by the NV center
should be the same in the bulk and in a nanocrystal, so that
local field corrections cancel out, and the
simple quantum electrodynamical analysis of eq. (\ref{na}) is valid.
We note that NV defects located very close to the nanocrystal surface would
have
different optical properties  than true NV centers
 owing
to the modification of local symmetry.
In particular their spectrum is expected to be modified, 
which is not what we have observed.

In order to evaluate the value of $n$ that appears in  eq. (\ref{na}),
we consider that (outside the local field area)
the NV center in bulk diamond  emits within an medium of index $n_d=2.4$.
For  the center in a sub-wavelength nanocrystal everything happens as if it
were
emitting in air for one half of the space,
and in fused silica (refractive index $n_s=1.45$) for the other half.
By using eq.(\ref{na}) in each half space we obtain
$1/\tau_{nc} = (1/2)[1/ (n_d \tau_b) + n_s/( n_d \tau_b)]$, giving
$\tau_{nc}  = 22.7$ ns
in good agreement with the experimental values.
We note that this value is independent from the nanocrystal size, provided
that this size is much smaller than the optical wavelength.
On the other hand, by looking at $10$ different nanocrystals,
we have found a dispersion of $\pm 4$ ns in the lifetime values.
This dispersion may be attributed to the position of the NV center
relative to the polymer/air interface, and to the randomness of
the dipole orientation \cite{LuK}.

As a conclusion, we have observed almost  background-free photon
antibunching from
single NV centers in diamond  nanocrystals at room temperature.
The photostability of NV centers in bulk diamond is preserved, allowing us
to lock
the laser beam on a single center during several hours.
Manipulation of nanocrystals is  a lot more flexible than bulk crystals.
Straightforward improvement of the light collection efficiency
should be possible by letting the nanocrystal sit on a
metallic mirror, or inserting it in a microcavity \cite{LK,GG,AK}.
It should be stressed that in bulk diamond individual NV centers can not be
observed close to the surface, owing to excessive stray light \cite{KMZW,ol}.
Nanocrystals have thus a clear advantage for inserting NV centers in
microcavities.
 These results show that diamond nanocrystals offer all the required
properties for
the realization of efficient single photon sources for quantum information
systems.
In addition, we found conclusive evidence that the lifetime
of a NV center is larger in nanocrystals than in bulk, owing to change
of the surrounding refractive index. Our data supports the conclusion
that the local field effects do not contribute to the change in lifetime.
This lifetime modification may then be interpreted as a simple quantum 
electrodynamics effect.

We thank E. Br\'eelle  from  the ``Groupe de Physique des Solides''
at Paris 6 for the sample irradiation, and A. Machu for sample annealing.
This work is supported by the European IST/FET program
``Quantum Information Processing and Telecommunication'',
project number 1999-10243 ``S4P".



\end{document}